\documentclass[journalm]{IEEEtran}

\usepackage[font=footnotesize]{caption}
\usepackage{graphicx}
\usepackage{subfig}
\usepackage{amsmath}
\usepackage{cite}

\begin{document}

\title{Generation of femtosecond optical vortex beams in all-fiber mode-locked fiber laser using mode selective coupler}


\author{Teng~Wang, Feng~Wang, Fan~Shi, Fufei~Pang, Sujuan~Huang, Tingyun~Wang, and~Xianglong~Zeng
\thanks{Manuscript received XXXXXX; revised XXXXXX; accepted XXXXXX. Date of publication XXXXXX; date of current version XXXXXX. This work was supported in part by the National Natural Science Foundation of China (Grant No. 11274224).}
\thanks{The authors are with the Key Lab of Specialty Fiber Optics and Optical Access Network, Shanghai University, Shanghai 200072, China. (e-mail: zenglong@shu.edu.cn)}
\thanks{Color versions of one or more of the figures in this paper are available online at http://ieeexplore.ieee.org.}
\thanks{Digital Object Identifier XXXXXX}}

\markboth{JOURNAL OF LIGHTWAVE TECHNOLOGY,~Vol.~XX, No.~X, XXX~XXXX}%
{Shell \MakeLowercase{\textit{et al.}}: Bare Demo of IEEEtran.cls for IEEE Journals}



\maketitle

\begin{abstract}

  We experimentally demonstrated a high-order optical vortex pulsed laser based on a mode selective all-fiber fused coupler composed of a single-mode fiber (SMF) and a few-mode fiber (FMF). The fused SMF-FMF coupler inserted in the cavity not only acts as mode converter from LP$_{01}$ mode to LP$_{11}$ or LP$_{21}$ modes with a broadband width over 100 nm, but also directly delivers femtosecond vortex pulses out of the mode locked cavity. To the best of our knowledge, this is the first report on the generation of high-order pulse vortex beams in mode-locked fiber laser. The generated 140 femtosecond vortex beam has a spectral width of 67 nm centered at 1544 nm.

\end{abstract}


%
\IEEEpeerreviewmaketitle

\section{Introduction}

\IEEEPARstart{O}{ptical} vortex beams (OVBs), also called orbital angular momentum (OAM) beams, are spatially structured beams with helical phase front. Such beams are characterized by a topological charge (order), and are found to carry an OAM of $\ell\hbar$ per photon, where $\ell$ is referred as the topological charge and can take any integer value, $\hbar$ is the reduced Planck constant \cite{allen1992orbital}. In the analytic expression, this helical phase front is usually related to a phase term of exp($\emph{i}\ell\theta$) in the transverse plane, where $\theta$ refers to the azimuthal coordinate and $\ell$ is an integer counting the number of intertwined helices. Therefore $\ell$ can assume a positive, negative or zero value, corresponding to clockwise or counter-clockwise phase helices or a Gaussian beam, respectively \cite{allen1999iv}. As vortex beams have a phase singularity, they have a doughnut-shaped spatial profile with zero intensity at the center. Due to the doughnut spatial structure and OAM properties, optical vortex beams attract a lot of attention in view of their applications, including optical tweezers \cite{padgett2011tweezers}, particle trapping \cite{kawauchi2007calculation}, high precision micromachining \cite{allegre2012laser, hnatovsky2010materials}, quantum computation \cite{arnaut2000orbital}, data transmission \cite{wang2012terabit}, optical communications \cite{su2012demonstration, bozinovic2013terabit, willner2015optical} and material processing \cite{meier2007material, koyama2011power, toyoda2012using, toyoda2013transfer, takahashi2016picosecond}.

\begin{figure}[htbp]
\centering
\includegraphics[width=1\linewidth]{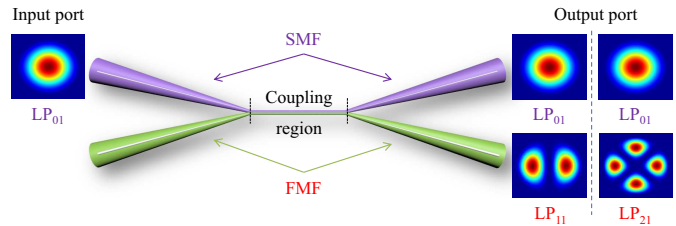}
\caption{Schematic of the MSCs, composed of a SMF and a FMF. Light is launched in the SMF input port; the LP$_{11}$ or LP$_{21}$ mode is expected to be preferentially excited at the FMF output port, while the uncoupled fundamental LP$_{01}$ mode will propagate along the SMF.}
\label{coupler}
\end{figure}

Driven by the distinctive properties and miscellaneous applications, there have been many attempts to generate optical vortex beams. Different methods to generate optical vortices both in free space and optical fibers have been proposed: spiral phase plates \cite{beijersbergen1994helical}, spatial light modulator (SLM) \cite{ostrovsky2013generation}, computer-generated holograms \cite{heckenberg1992generation}, cylindrical lens pairs \cite{beijersbergen1993astigmatic}, Q-plates \cite{marrucci2011spin, gregg2015Q-plates}, fiber gratings \cite{lin2014generation, zhang2016generation, zhao2016mode}, couplers \cite{yan2011fiber} and so on. All of these methods to generate optical vortices are investigated by using a continuous wave (CW), but many applications require OVBs with an ultrahigh peak power and narrow temporal pulse duration, such as in the field of material processing. Several attempts to generate pulsed OVBs have been reported. Recent years, Jianlang Li \emph{et al.} reported a radially polarized and passively Q-switched Yb-doped fiber laser. By using a Cr$^{4+}$:YAG crystal as a saturable absorber and a photonic crystal grating as a polarization mirror, a radially polarized pulse is produced \cite{li2008radially, lin2010efficient, lin2010radially}. but the scheme is not an all-fiber system which is not conducive to fiber system integration. Jiangli Dong \emph{et al.} demonstrated a passively mode-locked fiber laser that incorporates a two-mode fiber Bragg grating (FBG) for transverse-mode selection, where a mode-locked laser generates picosecond pulses at a fundamental repetition rate of 6.58 MHz for both the LP$_{01}$ and LP$_{11}$ modes \cite{dong2014mode}. However, it is known that the fiber grating only possesses narrow reflection bandwidth, which limits available spectrum of OVBs pulses. Thus using the fiber grating is difficult for achieving femtosecond pulse duration. In order to generate a laser pulse within femtosecond time domain, one needs to use a broad spectral bandwidth of mode conversion. Mode selective coupler (MSC) can work well at a broadband width. Therefore generating femtosecond vortex pulse is preferred based on fiber laser using a mode selective coupler.

In this paper, we present an all-fiber mode-locked fiber laser (MLFL) with femtosecond OVB pulse based on nonlinear polarization rotation (NPR). A fused SMF-FMF coupler is tactfully exploited to deliver pulse energy out of the fiber cavity, at the same time to achieve the mode conversion from LP$_{01}$ mode to LP$_{11}$ or LP$_{21}$ modes within a wide bandwidth over 100 nm. The high-order optical vortex pulse can be obtained by using a polarization controller (PC) on the FMF output port.
%

%

\section{Coupler fabrication and working principle}
Optical fiber couplers have a wide range of applications in optical fiber communication systems. Both single-mode fiber (SMF) and few-mode fiber (FMF) couplers are commercially available to be used as optical splitters/combiners. The proposed SMF-FMF coupler can not only be used as a splitter, but also as a mode converter.

\subsection{SMF-FMF coupler simulation}

Here, we present the schematic of SMF-FMF coupler which can excite high-order modes as shown in Fig. 1. The principle of the coupler is to phase match the fundamental mode in a single-mode fiber with a high-order mode in a few-mode fiber, and achieve mode conversion to high-order modes \cite{ismaeel2014all}. The larger the difference between these propagating constants, the lower the maximum power coupling efficiency. It is well known that the phase-matching condition can be achieved by satisfying the propagating constants of LP$_{01}$ mode in the SMF with LP$_{11}$ or LP$_{21}$ mode in the FMF according to the coupled-mode theory \cite{park2016broadband}. The high-order modes propagating in the FMF can be changed by pre-tapering the fiber diameters. Firstly, we need to understand the relationship between the mode effective indices and fiber diameters.

\begin{figure}[htbp]
\centering
\includegraphics[width=\linewidth]{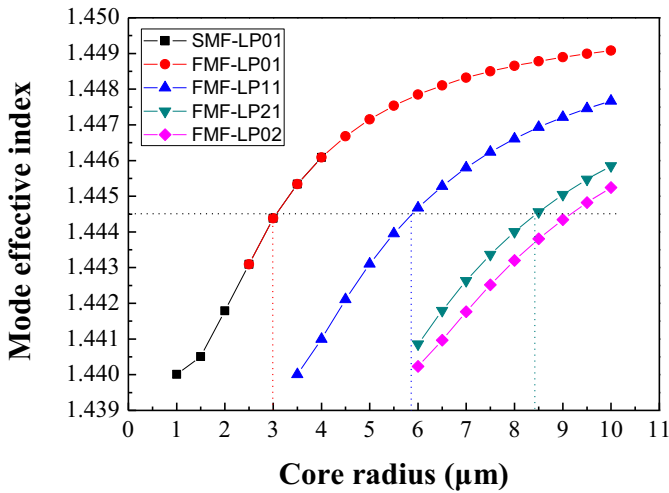}
\caption{The mode effective index curves for the LP$_{01}$ in the SMF and the desired high-order modes in the FMF versus core radius at the wavelength of 1550 nm.}
\label{Phasematch}
\end{figure}

The mode effective indices are calculated with the different fiber diameters for step-index fiber profiles in order to find the optimum diameter, at which the LP$_{01}$ modes are phase matched to the desired high-order modes. The phase-matching curves in Fig. 2 show that the thinner the core radius the lower the mode effective index. The dotted line represents the fiber dimensions chosen for the following simulations, which indicates that the SMF should be set to a core/cladding diameter of 6/93.75 $\mu$m, and FMF should be set to a core/cladding diameter of 11.6/72.5 $\mu$m, 16.9/105.6 $\mu$m, respectively, to meet phase-matching condition. The optimum taper diameters and the coupler length were optimized experimentally, until the desired mode and splitting ratio were obtained.

In order to determine the dependency of the coupling efficiency between the LP$_{01}$ mode and high-order modes on the fiber tapering diameter, we solved the following coupled equations:
\begin{equation}
\frac{dA_1(z)}{dz}=i(\beta_1+C_{11})A_1+iC_{12}A_2
\end{equation}
\begin{equation}
\frac{dA_2(z)}{dz}=i(\beta_2+C_{22})A_2+iC_{21}A_1
\end{equation}
where z represents the distance along the coupling region of the coupler, A$_{1}$ and A$_{2}$ are the slowly-varying field amplitudes in the SMF and FMF of the fused coupler, $\beta_1$ and $\beta_2$ are the propagation constants of LP$_{01}$ mode in the SMF and high-order mode in the FMF, respectively. Tapering the diameters of SMF and FMF changes their propagation constants. To make LP$_{01}$ mode in the SMF phase-matching with the desired high-order modes in the FMF, $\beta_{1}$ should be equal to $\beta_{2}$. $C_{11}$ and $C_{22}$, $C_{12}$ and $C_{21}$ are the self-coupling and mutual coupling coefficients, which are approximated as half the difference between the propagation constants of the symmetrical (even) and antisymmetrical (odd) modes on the composite waveguide. Self-coupling coefficients are small relative to mutual coupling coefficients, and can be ignore, also $C_{12}$ $\approx$ $C_{21}$ $\approx$ $C$, where $C$ depends on the width and length of coupling region. Thus, the power distributions in coupler are given by \cite{liu2015design}:
\begin{equation}
P_1(z)={\left| A_1(z) \right|}^{2}=1-F^{2}\sin^{2}(\frac{C}{F}z)
\end{equation}
\begin{equation}
P_2(z)=F^{2}\sin^{2}(\frac{C}{F}z)
\end{equation}
where $F=[1+\frac{{\beta_1-\beta_2}^{2}}{4C^{2}}]^{-1/2}$, $F^{2}$ is the maximum coupling power between the two fiber. According to Eq. (3) and (4), we can find that power in coupling region is exchange periodically. This suggests that by choosing a suitable interaction length, any arbitrary power distribution between the two interacting waveguides can be achieved. When propagation distance equal to coupling length, the power of LP$_{01}$ mode in SMF can be completely transferred to a certain higher-mode in FMF which meets the phase matching condition.

\begin{figure}[htbp]
\centering
\includegraphics[width=\linewidth]{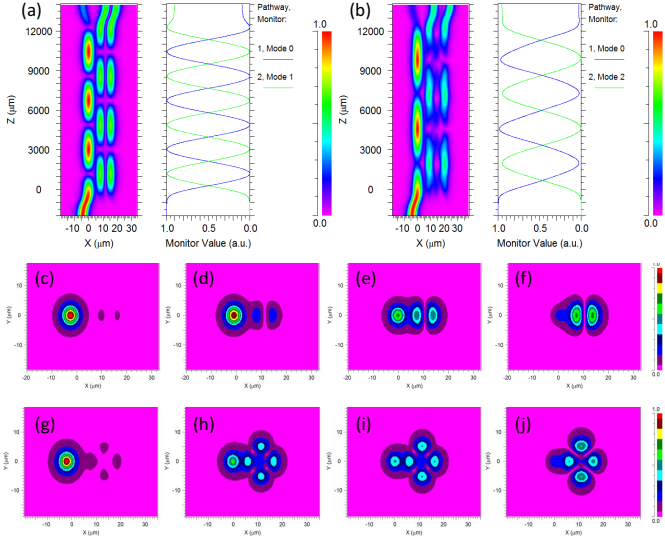}
\caption{Simulation results: (a) and (b): The power exchange in the coupling region when LP$_{01}$ mode in SMF converts to LP$_{11}$ or LP$_{21}$ mode in FMF; (c-f) and (g-j): The evolutions of mode field distribution in the coupling region when LP$_{01}$ mode converts to LP$_{11}$ or LP$_{21}$ mode in one period.}
\label{simulationresults}
\end{figure}

We use the commercial simulation software (Rsoft) to solve the modes propagating in the SMF-FMF couplers numerically and confirm the phase-matching condition as shown in Fig. 2. In our simulations, the SMF core radius is set to 3 $\mu$m, the FMF core radius is set to 5.8 $\mu$m, 8.45 $\mu$m, respectively, and the mode effective index is 1.4445. And the coupling region length is set to 12000 $\mu$m, the core distance between two fibers is set to 3 $\mu$m. From the simulation results, we can clearly see that the power in coupling region is exchange periodically. Fig. 3(a) and (b) show the power exchange in the coupling region when LP$_{01}$ mode in the SMF converts to LP$_{11}$ or LP$_{21}$ modes in the FMF. Fig. 3(c-f) show the mode field distributions along the coupling region when LP$_{01}$ mode converts to LP$_{11}$ mode in one period. Fig. 3(g-j) show the mode field distributions when LP$_{01}$ mode converts to LP$_{21}$ mode in one period. The evolutions of mode field distribution in the coupling region clearly present efficient mode conversion from LP$_{01}$ to desired high-order modes gradually.

\subsection{Mode conversion in SMF-FMF coupler}

A fused SMF-FMF fiber coupler basically consists of two parallel optical fibers that have been twisted, stretched and fused together  using oxyhydrogen flame so that their fiber cores are very close to each other. This forms a coupling region as shown in Fig. 1. The optical power couples from one fiber core to the other periodically in coupling region. The length of this coupling region determines the coupling ratio \cite{ismaeel2016removing}. Proper tapered fiber diameters are calculated by using numerical simulations of the SMF modes first. The SMF was tapered approximately to the target diameter where \emph{n}$_{eff}$ of 1550 nm became identical with that of the desired FMF high-order modes. Optimum taper diameters are determined after experimental fine tuning.  According to our simulation and experiment results, both SMF (core/cladding diameter = 8/125 $\mu$m) and FMF (core/cladding diameter = 20/125 $\mu$m) are not pre-tapered to obtain the phase-matching condition between LP$_{01}$ and LP$_{11}$, but for mode conversion from LP$_{01}$ to LP$_{21}$, pre-tapering SMF is required, because mode effective index of LP$_{21}$ is lower than LP$_{11}$ at the same diameter of FMF. Therefore tapering the SMF into a specific diameter will match the propagation constants for the LP$_{01}$ mode in the SMF with the LP$_{21}$ mode in the FMF.

 During the fusing process, light is launched into the input port, and the output powers from the output ports are carefully monitored. When the desired coupling ratio is achieved, the fully automated fusing process is stopped. The power is redistributed in the coupler region, and LP$_{11}$ mode or LP$_{21}$ mode is able to dominant in the output of FMF due to the phase-matching condition. Lastly, we encapsulate the fused SMF-FMF coupler using epoxy resin adhesive and heat shrink tubing. The excess loss of LP$_{11}$ and LP$_{21}$ SMF-FMF couplers are measured to be 0.9 dB, 2.7 dB, respectively.

\begin{figure}[!t]
\centering
\includegraphics[width=\linewidth]{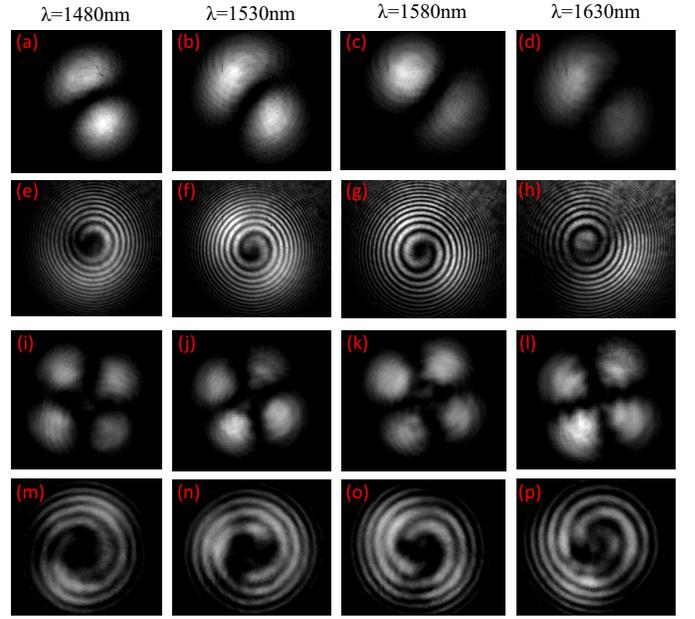}
\caption{CCD images of the output beam from mode selective couplers at wavelength of 1480 nm, 1530 nm, 1580 nm and 1630 nm. (a-d): Mode field distributions of LP$_{11}$ mode; (e-h): Spiral interferograms of corresponding optical vortices; (i-l): Mode field distributions of LP$_{21}$ mode; (m-p): Spiral interferograms of corresponding optical vortices.}
\label{Workingwidth}
\end{figure}

When the FMF is not deformed, the output beam is a high-order (LP$_{11}$ or LP$_{21}$) mode beam. Fiber OAM modes can be obtained by combining the degenerate odd and even LP$_{lm}$ modes with $\pm\pi/2$ phase difference (i.e., OAM$_{\pm l}$=LP$_{lma}\pm\emph{i}\times$LP$_{lmb}$), where \emph{l} refers to the azimuthal index and \emph{m} refers to the radial index. The LP$_{l>0m}$ modes have degenerate odd and even forms, which are sometimes labeled \emph{a} and \emph{b}. In order to generate OAM modes in weakly guiding fibers, production of a $\pm\pi/2$ phase difference between the odd and even LP$_{lm}$ mode becomes a key issue. McGloin \emph{et al.} proposed a simple method to generate the phase difference, i.e., stress the used fiber with rectangular lead weights \cite{mcgloin1998transfer}. A $\pm\pi/2$ phase difference could be obtained with the control of the fiber length and pressure. This is because the stress changes the effective dimensions of the FMF, leading to different phase velocities of the two modes. We use a polarization controller (PC) to press the FMF output port. By rotating the PC3 and adjusting the pressure appropriately, the two orthogonal modes can achieve a $\pi/2$ phase difference, and the output is an OAM beam with a circularly symmetric annular intensity profile and a helical phase front of exp($\pm\emph{i}\theta$).

 The output LP$_{11}$ and LP$_{21}$ mode patterns and their corresponding interference patterns from the FMF output port are observed at wavelength of 1480 nm, 1530 nm, 1580 nm and 1630 nm, as shown in Fig. 4. The output beam profiles are recorded by using a CCD camera (InGaAs camera, Model C10633-23 from Hamamtsu Photonics). From the near-filed beam profiles, we can confirm that the fused SMF-FMF coupler can be a all-fiber mode-converter with a broadband width over 100 nm from LP$_{01}$ mode to LP$_{11}$ or LP$_{21}$ modes.

\begin{figure}[!t]
\centering
\includegraphics[width=\linewidth]{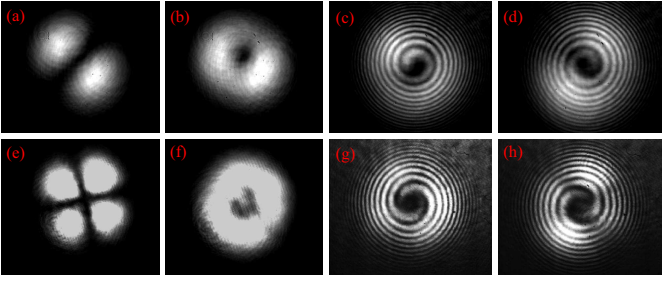}
\caption{CCD images of the output beam from mode selective couplers at the wavelength of 1550 nm. (a) and (e): intensity profiles of the LP$_{11}$ and LP$_{21}$ modes at the output FMF; (b) and (f): Donut-shaped mode patterns when pressing the output FMF; (c), (d), (g), and (h): corresponding spiral interferograms.}
\label{Modefield}
\end{figure}

The mode selections have a better performance at the communication wavelength of 1550 nm. As shown in Fig. 5(a) and (e), the experimental near-field patterns of LP$_{11}$ mode and LP$_{21}$ modes are successfully excited. Fig. 5(b) and (f) show the corresponding donut-shaped patterns when pressing the output FMF port. Fig. 5(c), (d), (g), and (h) show the interference patterns of the donut-shaped beams with a reference Gaussian beam, which indicate that the donut-shaped beams are vortex beams and the topological charge of vortices to be 1, 2, respectively. The clockwise spiral interference patterns for OAM$_{-1}$, OAM$_{-2}$ and the counterclockwise spiral interference patterns for OAM$_{+1}$, OAM$_{+2}$ clearly indicate that OAM$_{\pm1}$ and OAM$_{\pm2}$ are successfully achieved at the output FMF port.

\section{Experiments and discussion}
The schematic setup of a MLFL with ultrafast OVBs is shown in Fig. 6. The fused SMF-FMF fiber couplers are inserted in the mode-locked cavity of fiber laser to generate femtosecond pulses. The NPR mode-locking can be easily initialized by adjusting the PCs in the cavity \cite{deng200955}. The experimental near-field patterns of OVBs are successfully excited in output 1. The mode selective couplers are inserted into the fiber laser to excite LP$_{11}$ or LP$_{21}$ mode into the fiber laser, femtosecond OVBs with $\emph{l}=\pm$ 1 or 2 are obtained.

\begin{figure}[htbp]
\centering
\includegraphics[width=\linewidth]{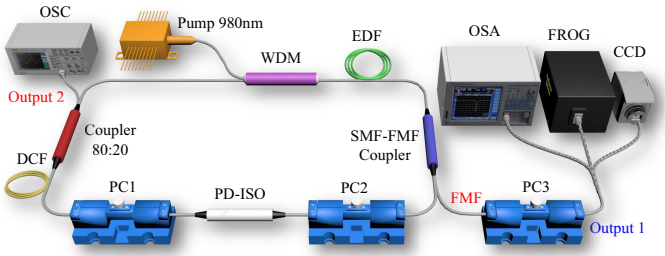}
\caption{Experimental setup used to excite OVB pulse. WDM: Wavelength Division Multiplexing coupler; EDF: Erbium-doped Fiber; DCF: Dispersion Compensating Fiber; PC: Polarization Controller; PD-ISO: Polarization-dependent Isolator; FROG: Frequency-resolved Optical Gating; OSC: Oscilloscope; OSA: Optical Spectrum Analyzer; CCD: Charge Coupled Device, Infrared camera.}
\label{Fiberlaser}
\end{figure}

 A length of 0.4 m heavily doped erbium-doped fiber (EDF) (LIEKKI, ER80-8/125) with group velocity dispersion of -20 ps$^{2}$/km is used in the experiment. The EDF with peak absorption of 80 dB/m at 1530 nm is acted as a gain medium. And it is pumped by a 980 nm laser diode with a maximum pump power of 620 mW through 980/1550 nm wavelength division multiplexing coupler (WDM). A length of 1.5 m dispersion compensating fiber (DCF) (THORLABS, DCF38) with group velocity dispersion of 50.35 ps$^{2}$/km is used to compensate negative dispersion fiber. A polarization-dependent isolator (PD-ISO) is inserted in the laser cavity to force the unidirectional laser operation and acts as the polarizer for the NPR mode-locking. The 80:20 optical coupler (OC) is used to extract the energy in the cavity for monitoring. The fused SMF-FMF coupler acts as power splitter and mode converter. Two polarization controllers (PCs) in the cavity are used to optimize the NPR. All the devices in the cavity are connected by standard SMF with GVD of -23 ps$^{2}$/km at 1550 nm. The total cavity length is 5.9 m, and the net cavity dispersion is 0.015 ps$^{2}$. The output spectrum is analyzed by an optical spectrum analyzer (YOKOGAWA, AQ6370C), and the time domain waveform is recorded by a 10 GHz electro-photonic detector (CONQUER, KG-PD-10G-FP) followed with a 1 GHz oscilloscope (Tektronix, MSO 4104). The signal is also sent to a commercial frequency-resolved optical gating (FROG, Mesa Photonics) for characterizing the pulse duration and the spectrum.

\begin{figure}[htbp]
\centering
\includegraphics[width=\linewidth]{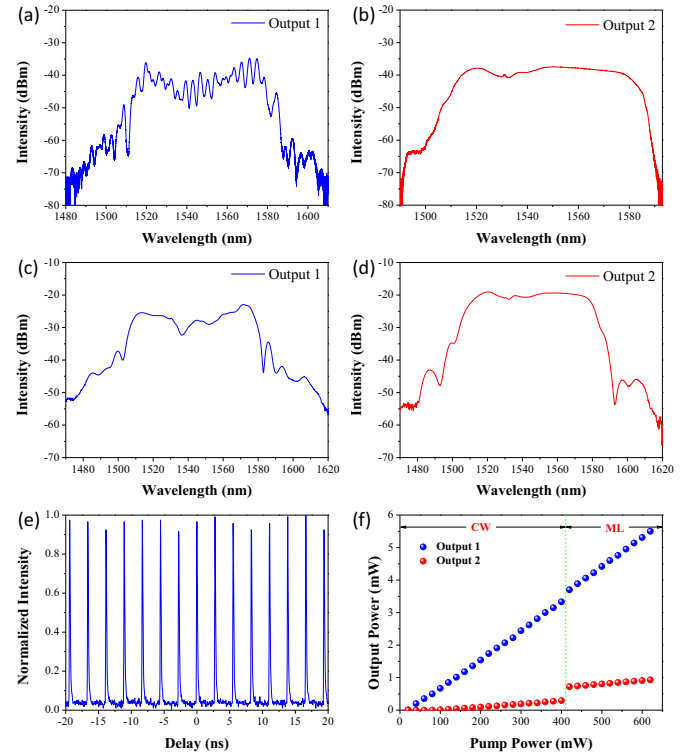}
\caption{(a) and (b): spectra of LP$_{11}$ mode in Output 1 at pump power of 620 mW and the corresponding spectrum at Output 2. (c) and (d): spectra of LP$_{21}$ mode in Output 1 at pump power of 620 mW and the corresponding spectrum at Output 2. (e): mode-locked pulse trains of femtosecond OAM$_{\pm2}$ mode. (f): the output powers from Output 1 and Output 2 as the function of pump power when mode selective coupler excited OAM$_{\pm2}$ mode.}
\label{Output}
\end{figure}

Firstly, we use an oscilloscope and optical spectrum analyzer (OSA) to observe the pulse trains and analyze the output spectrum, as shown in Fig. 7. The working state of fiber laser is monitored by using the signal of Output 2. Output 1 is analyzed under a pump power of 620 mW. Fig. 7(a) and (b) show the output spectra when the mode selective coupler excits LP$_{11}$ mode. A typical broad bandwidth spectrum is observed in Fig. 7(a) due to the positive net dispersion in the cavity \cite{tamura199377}. The central wavelength locates at 1547.4 nm, and the 3-dB bandwidth is measured to be 56.5 nm. The interference fringes in the FMF port are caused by the interference between the degenerated vector modes of LP$_{11}$ mode. LP$_{11}$ mode is degenerated by four vector modes possessing similar propagation constants, including TE$_{01}$, TM$_{01}$, HE$_{21}^{odd}$ and HE$_{21}^{even}$. Fig. 7(c) and (d) show the spectra of Output 1 and 2 when the mode selective coupler excites LP$_{21}$ mode. The central wavelength locates at 1545 nm, and the 3-dB bandwidth is measured to be 67.6 nm at Output 1. Fig. 7(e) presents the pulse trains of Output 1 with repetition rate of 36.10 MHz when mode selective coupler excited OAM$_{\pm2}$ mode. Fig. 7(f) shows the evolution of the output power versus the pump power for OAM$_{\pm2}$ mode. It is noticed that fiber laser is working at the CW state when the pump power is below 420 mW. Increasing the pump power continuously, self-starting and stable mode-locking is able to operate.

\begin{figure}[!t]
\centering
\includegraphics[width=\linewidth]{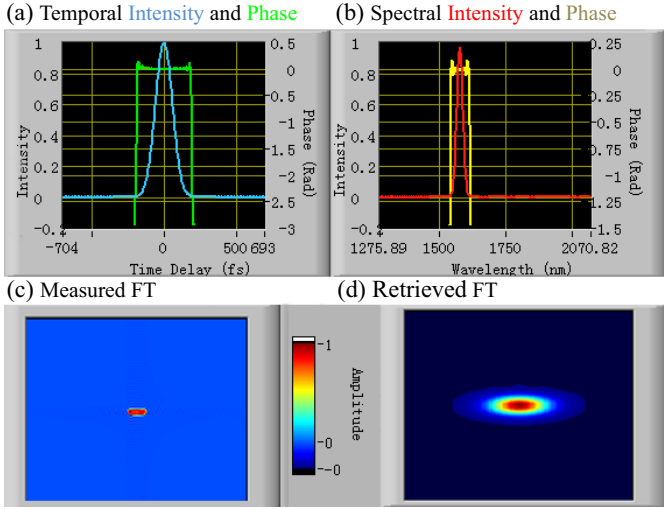}
\caption{FORG results of vortex pulses with OAM$_{\pm2}$: (a) Temporal Intensity and Phase profile, (b) Spectral Intensity and Phase profile, (c) Measured FROG traces, (d) Retrieved FROG traces.}
\label{FROG}
\end{figure}

Secondly, we use FROG to measure the pulse duration of OVBs with OAM$_{\pm1}$ and OAM$_{\pm2}$. The temporal and spectral intensities, phase profiles, frequency-doubled second harmonic generation (SHG) signal, the measured and retrieved FROG traces of the pulse from Output 1 are illustrated in Fig. 8. The pulse durations of OAM$_{\pm1}$ and OAM$_{\pm2}$ beams are measured to be 273 fs, 140 fs, respectively. And the time bandwidth product both are 0.45 from the retrieved signal.

\begin{figure}[!t]
\centering
\includegraphics[width=\linewidth]{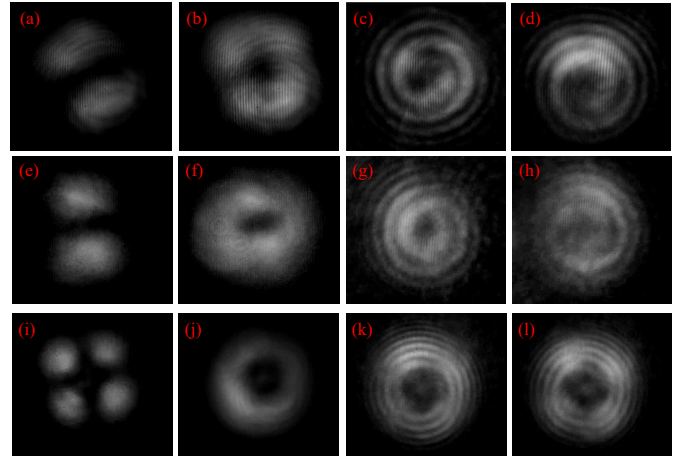}
\caption{CCD images of near-field intensity distribution of Output 1. (a), (e), and (i): intensity profiles of the LP$_{11}$ and LP$_{21}$ modes at the output FMF; (b), (f), and (j): donut-shaped mode patterns when pressing the output FMF; (c), (d), (g), (h), (k), and (l): corresponding spiral interferograms. The first row shows the near-field intensity distributions when the fiber laser is running at the state of CW. Both second and third row show the near-field intensity distributions when the fiber laser is running at the state of ML.}
\label{Vortexpulse}
\end{figure}

The fused SMF-FMF coupler is verified effectively when acting as a mode converter. The mode patterns from Output 1 are observed when the fiber laser is running at the state of CW and ML. As shown in Fig. 9, the experimental near-field patterns of LP$_{11}$ and LP$_{21}$ modes and the corresponding OAM$_{\pm 1}$ and OAM$_{\pm 2}$ modes are successfully obtained. Fig. 9(a-d) show the near-field intensity distributions when the fiber laser is running at the state of CW. Fig. 9(a) and (b) are the experimental near-field patterns of LP$_{11}$ mode and donut-shaped mode pattern when adjusting the PC3 on the FMF. Fig. 9(c) and (d) are the corresponding spiral interferogram. Fig. 9(e-l) show the near-field intensity distributions when the fiber laser is running at the state of ML. Fig. 9(e) and (i) show the experimental near-field patterns of LP$_{11}$ and LP$_{21}$ modes are successfully excited. Donut-shaped mode patterns of LP$_{11}$ and LP$_{21}$ modes are shown in Fig. 9(f) and (j). Efficient mode coupling occurs and different vector modes with polarization state are easily excited by adjusting the PC3 based on the principle of extrusion and torsion \cite{li2015controllable}. The intra-cavity PCs are carefully adjusted to obtain ideal OVBs for ensuring the mode-locking operation. Fig. 9(g), (h), (k), and (l) show the interference patterns of the donut-shaped beams with Output 2 (as a reference Gaussian beam), which indicates that femtosecond donut-shaped beams are vortex beams and the topological charge of vortices to be 1, 2, respectively. The clockwise spiral interference patterns for OAM$_{-1}$, OAM$_{-2}$ and the counter-clockwise spiral interference patterns for OAM$_{+1}$, OAM$_{+2}$ can be clearly seen, which shows that femtosecond OAM$_{\pm1}$ and OAM$_{\pm2}$ are successfully achieved at the output FMF port.

\section{Conclusion}
In conclusion, we show that mode selective couplers can be used to generate high-order OVBs and  mode couplings is analyzed in the coupling region between the fundamental mode in SMF and a high-order mode in FMF. We experimentally demonstrate an all-fiber optical vortex MLFL based on the mode selective coupler, and femtosecond optical vortex beams with topological charge of 1 and 2 are obtained from FMF output port of mode selective couplers, respectively. To the best of our knowledge, this is the first time to demonstrate such ultrafast MLFL supporting the generation of OVBs pulse by using the mode selective coupler. The measured time duration of OVB pulses are 273 fs, 140 fs of OAM$_{\pm1}$ and OAM$_{\pm2}$. A shorter pulse of the MLFL is supposed to be implemented by optimizing the dispersion of the laser cavity.


%


\section*{Acknowledgment}
This work was supported in part by the National Natural Science Foundation of China (Grant No. 11274224, 61635006). X Zeng acknowledges the Program for Professor of Special Appointment (Eastern Scholar) at Shanghai Institutions of Higher Learning and Science and Technology Commission of Shanghai Municipality (16520720900).

\ifCLASSOPTIONcaptionsoff
  \newpage
\fi



\bibliographystyle{IEEEtran}
\bibliography{References}

\end{document}